# Impact of Major RF Impairments on mm-wave Communications using OFDM Waveforms

Yaning Zou[1], Per Zetterberg [2], Ulf Gustavsson[3], Tommy Svensson[4], Ali Zaidi[3], Tobias Kadur[1], Wolfgang Rave[1], and Gerhard Fettweis[1]

[1] Vodafone Chair Technische Universität Dresden, Dresden, GERMANY
[2] Qamcom Research AB, Stockholm, SWEDEN
[3]Ericsson Research AB, SWEDEN
[4] Chalmers University of Technology, Gothenburg, SWEDEN
Contact emails: {yaning.zou, rave, fettweis}@ifn.et.tu-dresden.de, per.zetterberg@qamcom.se, {ulf.gustavsson, ali.zaidi}@ericsson.com, tommy.svensson@chalmers.se

*Abstract* —In this paper, we study the joint impact of three major RF impairments, namely, oscillator phase noise, power amplifier nonlinearity and I/Q imbalance on the performance of a mm-wave communication link based on OFDM modulation. General impairment models are first derived for describing the joint effects in each TX, each RX as well as a mm-wave communication link. Based on the obtained signal models and initial air interface design from the mmMAGIC project, we numerically evaluate the impact of RF impairments on channel estimation in terms of channel-to-noise ratio (CNR) and also channel fluctuation due to common phase error (CPE) caused by phase noise within the channel coherence time. Then the impact on the link performance in terms of maximum sum rate is evaluated using extensive computer simulations. The simulation results show that the used air interface design is generally robust to the presence of RF impairments. With regard to the use of high order modulation alphabet and implementation of low-power and low-cost RF transceivers in mm-wave communication, special attention needs to be paid on phase noise where the inter-carrier-interference (ICI) can become a major limiting factor.

## I. INTRODUCTION

In recent years, there has been intensive interests on the development of broadband communication service over mm-wave frequency bands. This has been considered by the telecommunication industry as a part of 5$^{th}$ generation (5G) multi-radio access technology (RAT) system architecture [1]-[3]. However, many challenges still need to be addressed for building a commercially viable network. One important issue in this context is the design of RF transceivers that operate at mm-wave frequency bands [4].

As discussed in [5], due to the limited transistor speeds and the limited supply voltage, transceiver design at the mm-wave range can be very challenging. In particular, it imposes huge difficulties on local oscillator (LO) design regarding to, e.g., LO generation, LO frequency division and LO distribution. At the same time, there are large constraints for producing output power at these frequencies due to material limitations in terms of charge carrier saturation velocity and the electric breakdown field [6]. Further on, components and manufacturing methods so far are also less refined at mm-wave frequencies, compared to traditional carrier frequencies for mobile communications. This further implies that the developed wireless equipment very often may be associated with strong impairments in the actual implementation. As a result, considering fundamental trade-offs over performance, power consumption and cost, the detrimental effects of RF impairments will be more pronounced in the transceivers operating at mm-wave frequencies. This could be even a bigger problem if a spectrally more efficient modulation scheme, e.g., OFDM modulation, is deployed. In order to avoid complication and risks during the deployment stage, it is necessary to take RF impairments into account already at the air interface design stage [7]. Meanwhile, with a proper air interface design, digital compensation algorithms can also be applied to mitigate negative effects. However, this requires a reference signal design that enables the use of efficient compensation algorithms.

In this paper, we study the impact of three major RF impairments in a direct-conversion transceiver, i.e., local oscillator phase noise, power amplifier (PA) nonlinearities and transceiver I/Q imbalances, on the link performance of mm-wave communications using OFDM waveforms. First, signal models for modeling individual RF impairments are discussed separately. Then incorporating different RF impairments, general impairment models are derived for describing the joint effects in each transmitter (TX), each receiver (RX) as well as the whole mm-wave communication link. Based on the obtained signal models and example system setups, we numerically evaluate the impact of RF impairments on channel estimation in terms of channel-to-noise ratio (CNR) and also error vector magnitude (EVM) of channel fluctuations due to common phase error (CPE) caused by phase noise. Then the impact on the link performance is also evaluated in terms of maximum sum rate using extensive computer simulations. The simulation results generally show that, out of the three considered RF impairments, inter-carrier interference (ICI) stemming from phase noise is the main performance degradation factor. It should be carefully taken into account in the physical layer design.

## II. Major RF Impairments in mm-wave Direct-Conversion Transceivers

As a starting point, we briefly introduce signal models of three major RF impairments in a direct-conversion transceiver, namely, local oscillator phase noise, PA nonlinearity, I/Q imbalance. We assume the ideal baseband signal in each model to be $N$ subcarriers OFDM modulated waveform $x(t)$. In the frequency domain, the input signal at the $k$-th subcarrier is denoted as $X_k$.

### A. Phase Noise

Local oscillators generally consist of a reference oscillator (or clock), a voltage controlled oscillator (VCO), and a phase-locked loop (PLL) with frequency divider, phase-frequency detector, charge pump and loop filter, see e.g. [15]. The noise in the active and lossy elements causes jitter in the oscillation period, which manifests itself as phase noise. Mathematically we model this effect as

$$x_{PN,T}(t) = x(t) e^{j(2\pi f_c t + \phi_{PN,T}(t))} \\ x_{PN,R}(t) = x(t) e^{-j(2\pi f_c t + \phi_{PN,R}(t))}, \quad (1)$$

where $\phi_{PN,T}(t)$ and $\phi_{PN,R}(t)$ refer to the oscillator phase noise at the TX and RX respectively. In OFDM systems this spectrum widening manifests itself as CPE rotation of all the subcarrier symbols within an OFDM symbol and ICI between the subcarriers. The shape of the phase-noise spectrum affects the distribution of the inter-carrier interference as a function of subcarrier spacing. This can been seen in a frequency domain model where the baseband equivalent signal at the subcarrier $k$ ($k = 0,\ldots,N-1$) reads [8]

$$X_{PN,T(R),k} = X_k J_{0,T(R)} + \sum_{l=0, l\neq k}^{N-1} X_l J_{k-l,T(R)} \\ = X_k J_{0,T(R)} + \xi_{PN,T(R),k}, \quad (2)$$

where $J_{i,T(R)} = \sum_{n=0}^{N-1} e^{j\phi_{PN,T(R)}(t)} e^{-j2\pi ni/N} / N$ and $\xi_{PN,T(R),k} = \sum_{l=0,l\neq k}^{N-1} X_l J_{k-l,T(R)}$. The spectrum of the phasor $e^{j\phi_{PN,T(R)}(t)}$ is known as the phase noise spectrum. As shown in [8], we can generally approximate $J_{0,T(R)} \approx e^{j\phi_{CPE,T(R)}}$ where $\phi_{CPE,T(R)}$ is the so-called CPE and corresponds to the mean rotation during one OFDM symbol. The additive ICI term $\xi_{PN,T(R),k}$ changes with different signal input and is generally not a complex Gaussian distributed random variable. The level of phase noise generally increases linearly with increased operating frequency.

### B. Power Amplifier

According to early work by Johnson, [6], due to physical constraints dictated by the breakdown field and the maximum carrier drift velocity, the output power capability scales inversely to the squared centre-frequency for any given transistor technology. This implies difficulties in producing high output power at mm-wave frequencies. The most common way of circumventing this issue, is to distribute the power amplification over the array by using smaller power amplifiers. This approach, however, requires accurate phase alignment and mutual coupling calibrations among all the used amplifiers at the antennas to enable effective beamforming. The sensitivity toward calibration errors is however reduced as the number of antennas grows large [9].

This fundamental trade-off will play an important role on the achievable link and system performance. Considering the fact that the practical PA power efficiency at mm-wave frequencies is much less than its counterpart operating at cm-wave frequencies due to intrinsic loss mechanisms, see e.g. [10], it can be expected that mm-wave PA may most likely have to work in a nonlinear region during transmission.

In this paper, we will represent the nonlinear behavior of the PA through a memoryless nonlinear function $f(\bullet)$. Based on Bussgang's theorem, [12], the statistical linear approximation of the input-to-output relation is described by

$$x_{NL}(t) = f(x(t)) = \alpha_1' x(t) + \xi_{NL}(t), \quad (3)$$

where $\alpha_1' = E[x_{NL}(t) x^*(t)] / \sigma_x^2$ and $\xi_{NL}(t)$ refers to the residual zero-mean complex Gaussian distributed distortion noise with power of $\sigma_{\xi_{NL}}^2 = E[|x_{NL}(t) - \alpha_1' x(t)|^2]$. In frequency domain, the nonlinear signal model at the $k$-th subcarrier reads

$$X_{NL,k} = \alpha_1' X_k + \Xi_{NL,k}, \quad (4)$$

where $\Xi_{NL,k}$ refers to the residual distortion $\Xi_{NL}(t)$ at the frequency bin $k$. The linear term $\alpha_1'$ can be considered constant over the coherence interval as long as the average transmit power does not change. For the sake of simplicity, we only consider in-band distortion part of PA nonlinearity in this paper. For a detailed out-of-band investigation, please refer to [11].

### C. I/Q Imbalance

I/Q imbalance stems from the unavoidable amplitude and phase mismatching of the transceiver I and Q chains. There have been extensive studies in the literature on the analysis and compensation of I/Q imbalances for cm-wave radio, e.g., in [13] and references therein. In the mm-wave circuit design, the I/Q imbalance will still play a role if direct-conversion or low-IF topology is used in the transceiver design. As shown in [13], the frequency-selective TX (T) and RX (R) I/Q imbalance baseband equivalent models are given by

$$x_{IQ,T(R)}(t) = g_{1,T(R)}(t) * x(t) + g_{2,T(R)}(t) * x^*(t), \quad (5)$$

where $g_{1,T}(t) = (\delta(t) + h_T(t) g_T e^{j\phi_T})/2$, $g_{2,T}(t) = (\delta(t) - h_T(t) g_T e^{j\phi_T})/2$, $g_{1,R}(t) = (\delta(t) + h_R(t) g_R e^{-j\phi_R})/2$ and $g_{2,R}(t) = (\delta(t) - h_R(t) g_R e^{j\phi_R})/2$. Here, $\{g_T, \phi_T\}$ and $\{g_R, \phi_R\}$ represent the amplitude and phase imbalances of the TX and the RX quadrature mixing stages, respectively. The filters $h_T(t)$ and $h_R(t)$ represent the I and Q branch frequency-response differences, in the TX and RX, respectively, In frequency domain [13], I/Q imbalance generates mirror-frequency interference as

$$X_{IQ,T(R),k} = G_{1,T(R),k}X_k + G_{2,T(R),k}X^*_{N-k+1}, \quad (6)$$

where $G_{T(R),1,k}$ and $G_{T(R),2,k}$ are the FFT values of $g_{T(R),1}(t)$ and $g_{T(R),2}(t)$ at bin $k$. In general, $g_{1,T(R)}(t)$, $g_{2,T(R)}(t)$, $G_{1,T(R),k}$ and $G_{2,T(R),k}$ change very slowly compared to the channel coherence time and can be considered constant over a long period of time [13].

## III. GENERAL IMPAIRMENT SIGNAL MODELS UNDER MULTIPLE RF IMPAIRMENTS

In practice, all the considered RF impairments co-exist in the transceiver. Therefore, it is important to understand also the joint effects of multiple RF impairments in the TX, RX and a mm-wave communication link.

### A. General Impairment Model at the Transmitter

In the TX, we have oscillator phase noise, PA nonlinearity and I/Q imbalance. Assuming the signal at the TX input to be $s(t)$ ($S_k$ at the $k$-th subcarrier), by combing (1), (5) and (3) sequentially, the time-domain general impairment model at the TX can be written as

$$\begin{aligned}x_{TX}(t) = & a'_1 g_{1,T}(t) * s(t)e^{j\phi_{PN,T}(t)} \\ & + a'_1 g_{2,T}(t) * s^*(t)e^{j\phi_{PN,T}(t)} + \xi_{NL}(t).\end{aligned} \quad (7)$$

In the frequency domain, incorporating (2) with (4) and (6), we have the impaired signal at the $k$-th subcarrier as

$$\begin{aligned}X_{TX,k} = & a'_1 \left( G_{1,T,k}S_k J_{0,T} + G_{2,T,k}S^*_{N-k+1}J_{0,T} \right) \\ & + \xi_{ICI,T,k} + \Xi_{NL,k},\end{aligned} \quad (8)$$

where $\xi_{ICI,T,k} = a'_1 \left( G_{1,T,k}\xi_{PN,T,k} + G_{2,T,k}\xi_{PN,T,N-k+1} \right)$. In effect, the transmitted signal is impaired by the signal at the mirror-frequency subcarrier, general ICI term $\xi_{ICI,T,k}$ and the residual nonlinear distortion noise $\Xi_{NL,k}$.

### B. General Impairment Model at the Receiver

At the RX side, assuming implementation of a linear LNA, we consider the joint effects of phase noise and I/Q imbalance. Also, assuming the receiver input signal to be $z(t)$ ($Z_k$ at the $k$-th subcarrier), by incorporating time-domain impairment models (1) and (5), the general impairment model at the RX is obtained as

$$z_{RX}(t) = \left( g_{1,R}(t) * z(t)e^{j\phi_{PN,R}(t)} + g_{2,R}(t) * z^*(t)e^{-j\phi_{PN,R}(t)} \right). \quad (9)$$

In the frequency domain, the signal model at the $k$-th subcarrier appears to be

$$Z_{RX,k} = G_{1,R,k}J_{0,R}Z_k + G_{2,R,k}J^*_{0,R}Z^*_{N-k+1} + \xi_{ICI,R,k}, \quad (10)$$

where $\xi_{ICI,R,k} = G_{1,R,k}\xi_{PN,R,k} + G_{2,R,k}\xi^*_{PN,R,N-k+1}$. As shown in (10), the main distortion is stemming from mirror-frequency interference and ICI $\xi_{ICI,R,k}$.

### C. General Impairment Model for a Communication Link

Now we consider the joint effects of RF impairments in a mm-wave communication link that consists of one TX, one RX and an effective transmission channel. In general, the mm-wave channel model has been regarded as one of the most important topics studied in the literature, e.g., [4] and references therein. The focus in this paper is, however, to examine the RF impairment aspect of mm-wave communications. In this context, we only consider the effective channel that combines the propagation channel between the TX and RX antennas and analog beamformers at the TX and RX. Assuming only one RF chain is used at the TX and RX sides, we denote the effective channel as a SISO channel impulse response $h(t)$ and the $m$-th OFDM symbol at the TX output as $x_{TX,m}(t)$. After experiencing the effective channel $h(t)$, the corresponding reception at the RX input appears to be

$$z_m(t) = x_{TX,m}(t) * h(t) + n_m(t), \quad (11)$$

where $n_m(t)$ refers to zero-mean complex Gaussian distributed channel noise at the $m$-th OFDM symbol interval. Incorporating (11) with (7) and (9), the time domain link model under the considered RF impairments reads

$$\begin{aligned}z_{RX,m}(t) \approx & \left( a'_1 g_{1,T}(t) * s_m(t)e^{j\phi_{PN,T,m}(t)} * h(t) \right) e^{j\phi_{PN,R,m}(t)} * g_{1,R}(t) \\ & + \left( a'_1 g_{2,T}(t) * s^*_m(t)e^{j\phi_{PN,T,m}(t)} * h(t) \right) e^{j\phi_{PN,R,m}(t)} * g_{1,R}(t) \\ & + \left( a'_1 (g_{1,T}(t) * s_m(t)e^{j\phi_{PN,T,m}(t)} * h(t))^* \right) e^{-j\phi_{PN,R,m}(t)} * g_{2,R}(t) \\ & + \left( \xi_{NL,m}(t) \right) * h(t) e^{j\phi_{PN,R,m}(t)} * g_{1,R}(t) + g_{1,R}(t) * n_m(t)e^{j\phi_{PN,R,m}(t)},\end{aligned} \quad (12)$$

where the subscript $m$ in all terms denotes the corresponding function at the $m$-th OFDM interval. Assuming the cyclic prefix is longer than the overall delay spread of the effective channel $h(t)$, we can translate the time domain model in (12) into the frequency domain as

$$\begin{aligned}Z_{k,m} \approx & \bar{H}_{k,m}S_{k,m} + \vec{H}_{k,m}S^*_{N-k+1,m} + \xi_{ICI,R,k,m} \\ & + H_k J_{0,R,m} G_{1,R,k}(\xi_{ICI,T,k,m} + \Xi_{NL,k,m}) + n_k,\end{aligned} \quad (13)$$

where $\bar{H}_{k,m} = a'_1 J_{0,T,m}J_{0,R,m}G_{1,T,k}G_{1,R,k}H_k$ and $H_k$ is the channel frequency response at the $k$-th subcarrier $\vec{H}_{k,m} = a'_1 \times (J_{0,T,m}J_{0,R,m}G_{2,T,k}G_{1,R,k} + J_{0,T,m}J^*_{0,R,m}G^*_{1,T,N-k+1}G_{2,R,k})$.

Based on (13), in addition to the additive channel noise, the received signal can be divided into 4 parts.

1) The desired signal $\bar{H}_{k,m}S_{k,m}$. After experiencing linear gain of PA, I/Q imbalance as well as CPE rotation at TX and RX, the equivalent channel for the subcarrier $k$ has changed from the effective channel alone $H_k$ to be $\bar{H}_k$. In general, the effective channel, PA nonlinear behavior and I/Q imbalance parameters can be assumed to be unchanged if the transmission time is much smaller than the coherence time while the CPE term changes from one OFDM symbol to another due to the fact that the phase noise is a random process. The level of variation depends on phase noise parameters as well as the used moudulation and system setups. This issue will be studied in more details in Subsection IV. B.

2) Mirror-frequency interference $\vec{H}_{k,m}S^*_{N-k+1}$. This term is caused by the I/Q imbalance effect. As shown in [13], I/Q imbalance compensation can be carried out with a reasonable amount of resources using both blind and pilot-based algorithms.

3) ICI term due to phase noise in TX and RX $H_k J_{0,R} G_{1,R,k} \xi_{ICI,T,k} + \xi_{ICI,R,k}$. This term is generally a linear summation of all the subcarriers with linear combining coefficients that change in every OFDM symbol. The level of interference depends on the characteristics of phase noise and also the used subcarrier spacing in OFDM modulation.

4) Nonlinear distortion term $H_k J_{0,R,m} G_{1,R,k} \Xi_{NL,m,k}$. This term is caused by PA nonlinear behavior and also varies in different OFDM symbol durations.

## IV. IMPACT OF RF IMPAIRMENTS ON MM-WAVE COMMUNICATION LINK

In this section, we study the impact of all the considered RF impairments on the link performance with respect to two aspects. First, we evaluate the impact of RF impairments on the channel estimation quality. Second, we evaluate the performance degradation in terms of instantaneous signal-to-interference-plus-noise ratio (SINR) and achievable sum rate.

### A. Impact of RF Impairments on Channel Estimation

For channel estimation purposes, reference signal is typically used in the wireless communication system. Assuming a pilot symbol $S_{P,k}$ allocated to the $k$-th subcarrier, and incorporating the link level model for the $m$-th received OFDM symbol in (13), the corresponding reception at RX output reads

$$Z_{p,k,m} \approx \bar{H}_{k,m}S_{P,k} + \vec{H}_{k,m}S^*_{N-k+1,m} + \xi_{ICI,R,k,m} \\ + H_k J_{0,R,m} G_{1,R,k}(\xi_{ICI,T,k,m} + \Xi_{NL,k,m}) + n_k. \quad (14)$$

In case a pilot signal $S^*_{p,N-k+1,m}$ is also allocated to the mirror frequency $N-k+1$, the second term $\vec{H}_{k,m}S^*_{N-k+1}$ can be considered as a useful term if the channel estimation is carried out for each mirror-frequency pair. On the other hand, in case a data symbol is allocated at the mirror-frequency subcarrier $N-k+1$, the second term $\vec{H}_{k,m}S^*_{N-k+1}$ is generally an interference term unless more sophisticated processing is carried out.

As a concrete example, here we assume least-square (LS) estimation is applied for channel estimation of the equivalent transmission channel $\bar{H}_{k,m}$ as

$$\hat{\bar{H}}_{k,m} \approx Z_{p,k,m} / S_{p,k}. \quad (15)$$

The resulting channel estimation error can be quantified using the instantaneous channel to noise ratio (CNR) as

$$CNR_k = \frac{|\bar{H}_{k,m}|^2}{|\hat{\bar{H}}_{k,m} - \bar{H}_{k,m}|^2}. \quad (16)$$

In the wireless communication context, receiver additive noise is normally regarded as the main source for the channel estimation errors. However, with the presence of RF impairments in the mm-wave transmission link, an estimation error floor will be formed.

### B. Channel Flucutation due to CPE

In many wireless communication systems, e.g., LTE standards [14], in order to reduce system overhead, the reference signal for channel estimation is deployed sparsely over time domain. Then averaging or interpolation can be applied to obtain channel estimates in all the symbol intervals of, e.g., one subframe. This is a valid approach if the duration of the subframe is much shorter than the channel coherence time.

However, as shown in (12) and (13), taking the used transceiver chain into account, the overall equivalent channel at the $m$-th OFDM symbol interval becomes to be $\bar{H}_{k,m} = a'_1 J_{0,T,m} J_{0,R,m} G_{1,T,k} G_{1,R,k} H_k$. Within coherence time, $a'_1 G_{1,T,k} G_{1,R,k} H_k$ are generally unchanged when the TX and RX combined CPE term $J_{0,T,m} J_{0,R,m}$ varies over different symbol durations. If the system designer doesn't take this aspect into account and makes the assumption that the channel doesn't change within the coherence time based on $H_{k,m_1} = H_{k,m_2}$ ($m_1 \neq m_2$) and reduces system overhead by deploying reference signal only at the $m_1$-th OFDM symbol and acquiring the channel for the $m_2$-th OFDM symbol based on the channel for the $m_1$-th OFDM symbol, estimation errors are unavoidable even without any channel noise as we have $\bar{H}_{k,m_1} \neq \bar{H}_{k,m_2}$ in practice. The resulting EVM for the estimation appears to be

$$EVM = 10\log_{10}\left(|\hat{\bar{H}}_{k,m_2} - \bar{H}_{k,m_2}|^2 / |\bar{H}_{k,m_2}|^2\right). \quad (17)$$

### C. Impact of RF Impairments on Data Transmission

During data transmission phase, the interference caused by all the considered RF impairments can dramatically degrade link performance. Based on (13), with given propagation channel

$H_k$ and assuming perfect knowledge of $\bar{H}_k$, we define the instantaneous SINR at the $k$-th subcarrier as

$$\gamma_k = \frac{\left|\bar{H}_k^2\right|\sigma_s^2}{E\left[\left|Z_k - \bar{H}_k S_k\right|\right]}, \quad (18)$$

where the expectation is carried out over different symbol and phase noise realizations. Inherently, the average sum rate of the mm-wave link under multiple RF impairments reads

$$R_A \approx \frac{1}{N}\sum_{k=0}^{N-1} E\left\{\log_2(1+\gamma_k)\right\}. \quad (19)$$

### D. Air Interface Design with respect to the considered RF Impairments

The impact of RF impairments on system performance depends on the quality of the used transceivers and also on the air interface design. In general, the RF impairment problem is more pronounced in the high SNR region than in the low SNR region where additive channel noise is the dominant adverse factor. In mm-wave communications, the system most likely will operate at low and middle SNR regions due to strong path loss. This implies that, compared to its counterpart at lower carrier frequencies, the considered RF impairments will play a rather big role when the system design intends to push towards highly spectrum efficient transmission.

In this case, we provide some general insights for coping with the considered RF impairments from air interface design point of view as follows.

First of all, I/Q imbalance is normally the easiest impairment to cope with among the considered three impairments. The impairment parameters change very slowly over time and the resulting mirror-frequency interference could be mitigated using reasonable system resources, e.g., if pilot signal is available at the mirror-frequency subcarriers [13]. However, such a paired allocation is not supported in the current LTE standard [14] with, e.g., the demodulation reference signal (DM-RS).

Regarding the nonlinear in-band distortion generated by the PA, peak-to-average power ratio (PAPR) reduction or pre-distorter may potentially be deployed at the transmitter in order to improve the transmit signal quality. On the other hand, these tools may not be used to address the more critical issue of output power capability at mm-wave. In the end, using large antenna systems with distributed power amplification may form a more practical solution.

Last but foremost, the effects of phase noise should be carefully treated as its level scales up with operating frequencies. CPE tracking is in general rather straight-forward and yet requires allocating a small amount of pilot signal overhead in each OFDM symbol. On the other hand, combined with the effects of frequency-selective channel, ICI mitigation can be very challenging. By proper air interface design, e.g. using wide subcarrier spacing as proposed in [17], the negative impact of ICI on link performance can be alleviated. But, one still needs to be aware of the possible performance degradation if, e.g., a low-power and low-cost transceiver is implemented at the UE side and/or high order modulation alphabet is used as the subcarrier modulation alphabet in the OFDM waveform. In this case, in addition to continuously improved integrated circuit design at mm-wave frequency, devising efficient ICI cancellation schemes, e.g, that only mitigate major interference from neighboring subcarriers could form a reasonable solution. Such digital compensation approaches, however, require again a dedicated reference signal structure in every OFDM symbol as well as careful evaluation of symbol likelihoods and estimation of the phase noise process in an efficient manner.

## V. NUMERICAL RESULTS AND ANALYSIS

### A. System Setups

In this section, we examine the impact of the considered RF impairments using extensive computer simulations with the initial air interface design proposed by the mmMAGIC project [17]. Assume the system operates at a central frequency of 28 GHz or 82 GHz. The channel delay profile in [4] is implemented to model the multi-path effect of the propagation channel $h(t)$. The used OFDM waveform consists of 2048 subcarriers with 60 kHz subcarrier bandwidth at 28 GHz and 480 kHz subcarrier bandwidth at 82 GHz. The CP length is assumed to be 144 samples. The example subcarrier modulation is set to be 16 QAM. The array gain is assumed to be 30 dB. The SNR in the simulation is defined as received SNR plus array gain. As discussed in Section III, the mm-wave communication system is generally expected to operate at the low and middle SNR regions. Here, to characterize the general behavior of the OFDM system, we consider an SNR range from 0 to 35 dB.

Fig. 1 shows the spectrum of seven state-of-the-art local oscillators consuming below 50mW (except the one in solid black which consumes 560mW). Local oscillators with higher power consumption will generally have better performance since this allows increased signal to noise ratios inside the circuit, see e.g. [16]. For all the benchmarking local oscillators it is customary to normalize the spectrums level with the carrier frequency squared $f^2$. This has been done in Fig. 1 so that all spectrums are translated to a nominal 50GHz carrier frequency. In the mmMAGIC project, two phase noise models are proposed. The "low" phase noise model corresponds to the case that an oscillator with reasonable performance and quality is implemented whereas the "high" phase noise model considers the case that a low-cost and low-power transceiver is used and the oscillator quality is very poor. In the model, the phase-noise is generated by filtering Gaussian noise to represent the reference oscillator, the PLL components, the VCO and the loop filter. A Matlab script for generating data from the phase-noise model has been released as open source (the code will be released before publication of this paper). Two cases are considered in the simulations in this paper. Case I: good quality oscillators are used at the both TX and RX. Case II: good quality oscillator is used at the TX and low quality oscillator is used at the RX. The PA nonlinearity is modeled by the sum of $3^{rd}$, $5^{th}$, $7^{th}$, $9^{th}$ order memoryless polynomials. The frequency-selectivity I/Q imbalances are modeled with three-tap branch filters with phase and gain imbalance in the range of $1°$-$5°$ and 1%-5% respectively at the TX and RX corresponding to image rejection ratio (IRR) of 25-40 dB. For reference purpose, perfect I/Q match is also considered in the simulation in order to demonstrate the achievable performance with I/Q imbalance compensation.

*B. Simulation Examples*

First, the achievable CNR under all the considered RF impairments is evaluated numerically based on (16) for both Case I and Case II. As shown in Fig. 2, if the oscillator quality is reasonably good at the TX and RX, it is possible to achieve good channel estimation quality with the used air interface design. On the other hand, if phase noise is extremely strong as described in the "high" phase noise model case, a channel estimation quality upper bound of around 12 dB will be formed especially at 82 GHz.

Next, we examine the level of effective channel fluctuations stemming from varying CPEs. In details, the equivalent channels in two consecutive OFDM symbol durations are compared in terms of EVM as defined in (17). The resulting probability density functions (PDF) are illustrated in Fig. 4 over 10000 realizations. It clearly shows that CPE tracking is needed in each OFDM symbol with operating frequency at 82 GHz.

Based on (18) and (19), performance degradation due to RF impairments is demonstrated in terms of achievable sum rates. As shown in Fig. 6, considering SNR of 10-20 dB at 28 G Hz, the used air interface design is robust to the considered RF impairments even when the high phase noise model is used at the RX side. On the other hand, at 82 GHz, the phase noise can be problematic with the "high" phase noise model. Performance degradation is non-trivial even at the SNR of 15-20 dB.

At last, achievable SIRs due to ICI term with different subcarrier spacing at different carrier frequencies are compared using the "low" phase noise model proposed by the mmMAGIC project [17]. According to Fig. 7, with reasonably good transceiver implementations, it is possible achieve 35 dB SIR with subcarrier spacing of 60 kHz at 28 GHz and around 30 dB SIR with subcarrier spacing 480 kHz at 82 GHz. However, with the use of more cost and power efficient transceiver implementation, the achievable SIRs are expected to be lower. Digital or analog compensation techniques need to be developed for enabling spectrally efficient transmission schemes.

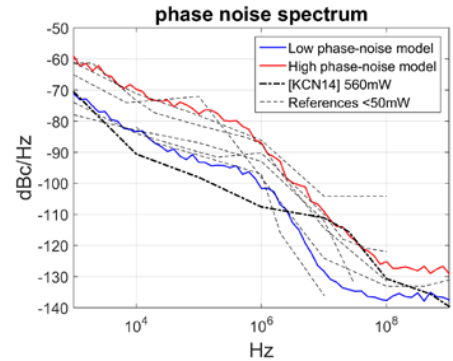

Fig. 1. Phase noise spectrum of seven local oscillators (thick dashed black line), phase noise model with "high" phase noise setting (red), and low phase-noise setting (blue).

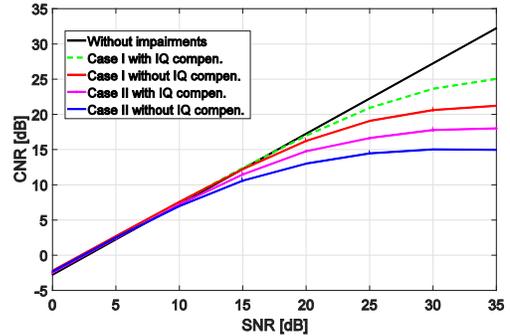

Fig. 2. Comparisons of CNR with and without RF impairments over 1000 realizations at the 28 GHz.

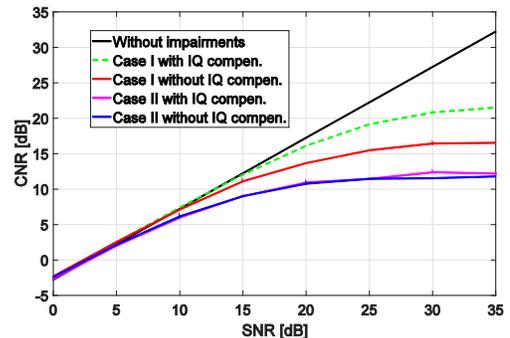

Fig. 3. Comparisons of CNR with and without RF impairments over 1000 realizations at 82 GHz.

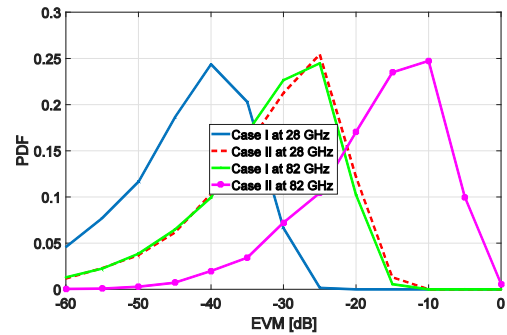

Fig. 4. PDF of difference on the equivalent channel between two consecutive OFDM symbols in terms of EVM as defined in (17) for both case I and case II at 28 GHz and 82 GHz.

## VI. CONCLUSIONS

In this paper, we studied the joint impact of three major RF impairments, namely, oscillator phase noise, PA nonlinearity and I/Q imbalance on the performance of a mm-wave communication link. Based on the developed general impairment model, the main negative effects are phase rotation on the equivalent channel, mirror-frequency interference, ICI interference and nonlinear distortion. The resulting performance degradation was evaluated in terms of both channel estimation quality and data transmission link quality. Based on simulation results, with reasonable good transceiver implementation, the air interface proposed by the mmMAGIC project is robust to the considered impairments. In case of highly spectrum efficient transmission and/or low-power and low-cost transceiver implementation, digital/analog compensation schemes need to be deployed.

## VII. ACKNOWLEDGEMENT

The research leading to these results received funding from the European Commission H2020 programme under grant agreement n°671650 (5G PPP mmMAGIC project).

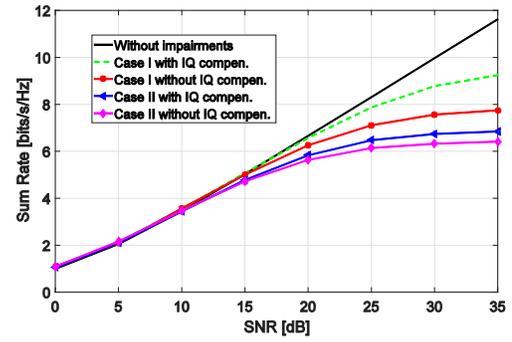

Fig. 5. Performance comparisons without and with RF impairments. Two oscillator implementations case I and case II are considered. Operating frequency is 28 GHz.

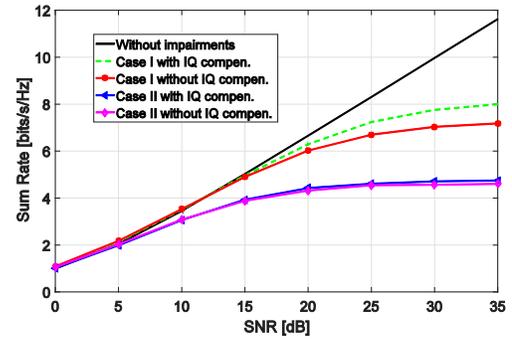

Fig. 6. Performance comparisons without and with RF impairments. Two oscillator implementations case I and case II are considered. Operating frequency is 82 GHz.

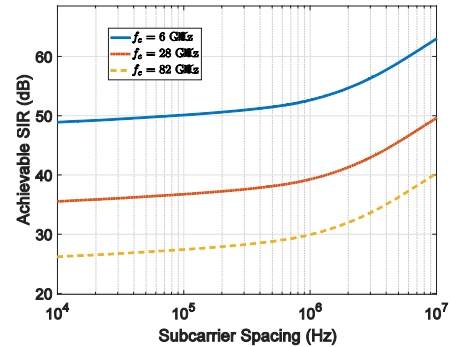

Fig. 7. Achievable SIRs due to ICI as a function of subcarrier spacing at different carrier frequencies, based on mmMAGIC "low" phase noise model.